\documentstyle[12pt]{article}
\evensidemargin\oddsidemargin
\def\Im{\mathop{{\cal I}\!m}}
\def\gev{{\rm\,GeV}}

\newcommand{\alt}{\mathrel{\raisebox{-.6ex}{$\stackrel{\textstyle<}{\sim}$}}}
\newcommand{\agt}{\mathrel{\raisebox{-.6ex}{$\stackrel{\textstyle>}{\sim}$}}}

\def\overlay#1#2{\ifmmode \setbox 0=\hbox {$#1$}\setbox 1=\hbox to\wd 0{\hss
$#2$\hss }\else \setbox 0=\hbox {#1}\setbox 1=\hbox to\wd 0{\hss #2\hss }\fi
#1\hskip -\wd 0\box 1}

\catcode`@=11
\newtoks\@stequation

\def\mathletters{\refstepcounter{equation}%
  \edef\@savedequation{\the\c@equation}%
  \@stequation=\expandafter{\theequation}
  \edef\@savedtheequation{\the\@stequation}
  \edef\oldtheequation{\theequation}%
  \setcounter{equation}{0}%
  \def\theequation{\oldtheequation\alph{equation}}}

\def\endmathletters{%
  \setcounter{equation}{\@savedequation}%
  \@stequation=\expandafter{\@savedtheequation}%
  \edef\theequation{\the\@stequation}%
  \global\@ignoretrue}
\catcode`=12

\begin{document}

\font\fortssbx=cmssbx10 scaled \magstep2
\hbox to \hsize{
\includegraphics{/NextLibrary/TeX/tex/inputs/uwlogo.ps}
\hskip.5in \raise.1in\hbox{\fortssbx University of Wisconsin - Madison}
\hfill\vbox{\hbox{\bf MAD/PH/781}
            \hbox{August 1993}} }

\baselineskip14pt

\begin{center}
{\Large\bf   SUPERSYMMETRY WITH GRAND\\[.1in]
 UNIFICATION\footnote{Talk presented by V.~Barger at the
{\it Workshop on Physics at Current Accelerators and the\break Supercollider},
Argonne, June 1993.}}\\[.2in]
{\large V.~Barger$^{\,a}$\llap,  M.S.~Berger$^{\,a}$\llap, P.~Ohmann$^{\,a}$
and R.J.N.~Phillips${\,^b}$}\\[.1in]
\it
$^a$Physics Department, University of Wisconsin, Madison, WI 53706, USA\\
$^b$Rutherford Appleton Laboratory, Chilton, Didcot, Oxon OX11 0QX, UK
\end{center}

\renewcommand{\LARGE}{\Large}
\renewcommand{\Huge}{\Large}

\section{Introduction}

   Supersymmetry (SUSY) has many well known attractions, especially in
the context of Grand Unified Theories (GUTs).  SUSY stabilizes scalar
mass corrections (the hierarchy problem), greatly reduces the number of
free parameters, facilitates gauge coupling unification, and provides a
plausible candidate for cosmological dark matter.  In this conference
report we survey some recent examples of progress in SUSY-GUT applications.

\section{Gauge coupling unification}

    As the renormalization mass scale $\mu$
is changed, the evolution of couplings is governed by the Renormalization Group
Equations (RGE).
For the gauge group $\rm SU(3)\times SU(2)\times U(1)$,
with corresponding gauge couplings
$g_3(=g_s), g_2(=g), g_1(=\sqrt{5/3}g')$, the RGE can be written
\begin{equation}
{dg_i\over dt} = {g_i\over 16\pi^2} \left[ b_ig_i^2 + {1\over16\pi^2}
\left( \sum_{j=1}^3 b_{ij} g_i^2g_j^2 - \sum_{j=1}^3 a_{ij}
g_i^2\lambda_j^2\right)\right] \,,
\end{equation}
where $t=ln( \mu /M_G)$ and $M_G$ is the GUT scale.
The first term on the right is the one-loop approximation; the second
and third terms contain two-loop effects, involving other gauge
couplings $g_j$ and Yukawa couplings~$\lambda_j$.  The coefficients $b_i,\
b_{ij}$ and $a_{ij}$ are determined at given scale $\mu$ by the content of
active particles  (those with mass ${}<\mu$).  If there are no thresholds
({\it i.e.} no
changes of particle content) between $\mu$ and $M_G$, then the coefficients
are constants through this range and the one-loop solution is
\begin{equation}
\alpha_i^{-1}(\mu)  =  \alpha_i^{-1}(M_G) - t b_i/(2\pi)   \;,
\end{equation}
where $\alpha_i = g_i^2/(4\pi)$; thus $\alpha_i^{-1}$ evolves linearly with
$\ln\mu$ at one-loop order.    If there are no new physics thresholds
between $\mu = M_Z \simeq m_t$  and $M_G$,  as in the basic Standard Model
(SM), then equations of this kind should evolve the observed couplings at the
electroweak scale~\cite{giatw}
\begin{eqnarray}
\alpha_1(M_Z)^{-1} &=& 58.89 \pm 0.11 \,, \\
\alpha_2(M_Z)^{-1} &=& 29.75 \pm 0.11 \,, \\
\alpha_3(M_Z) &=& 0.118\pm 0.007 \,,
\end{eqnarray}
to converge to a common value at some large scale.  Figure~1(a) shows that
such a SM extrapolation does NOT converge; this figure actually includes
two-loop effects but the evolution is still approximately linear versus
$\ln\mu$, as at one-loop order.  GUTs do not work, if we assume just SM
particles plus a desert up to $M_G$.

   But if we increase the particle content to the minimal SUSY model (MSSM),
with a threshold not too far above $M_Z$,  then
GUT-type convergence can happen.  Figure~1(b) shows an example with SUSY
threshold $M_{\rm SUSY}=1$~TeV~\cite{susygut,bbo}.  The evolved couplings are
consistent with a common
intersection at $M_G \sim 10^{16}$\,GeV; GUTs are plainly more successful with
MSSM than with SM. Henceforth we assume MSSM.  In fact a precise single-point
intersection is not strictly necessary; the exotic GUT gauge,
fermion and scalar particles may not be quite degenerate,
giving several non-degenerate thresholds near $M_G$, to be
passed through on the way to GUT unification.

\begin{center}
\hspace*{0in}

\parbox{5.5in}{\small Fig.~1. Gauge coupling evolution: (a)  in the SM;
(b)  in a SUSY-GUT example~\cite{bbo}.}
\end{center}

\section{Yukawa coupling evolution}

   The Yukawa couplings also evolve.  Typical evolution equations
are\cite{susyrge1,susyrge2}
\begin{mathletters}
\begin{eqnarray}
{d\lambda_t\over dt} &=& {\lambda_t\over16\pi^2} \left[-\sum c_i g_i^2 +
6\lambda_t^2 + \lambda_b^2  + \mbox{2-loop terms}\right] \;,\label{yuklam_t}\\
{d\lambda_b\over dt} &=& {\lambda_b\over16\pi^2} \left[-\sum c'_ig_i^2 +
\lambda_t^2 + 6\lambda_b^2 + \lambda_\tau^2 + \hbox{2-loop}\right]\,,\\
{d\lambda_\tau\over dt} &=& {\lambda_{\tau }
\over16\pi^2} \left[-\sum c''_ig_i^2 +
3\lambda_b^2 + 4\lambda_\tau^2 + \hbox{2-loop}\right]\,,
\end{eqnarray}
\end{mathletters}
with $c_i=(13/15,3,16/3)$, $c'_1=(7/15,3,16/3)$, $c''_i=(9/5,3,0)$, and hence
\begin{equation}
{d(\lambda_b/\lambda_\tau)\over dt} = {(\lambda_b/\lambda_\tau)\over16\pi^2}
\left[-\sum d_i g_i^2+\lambda_t^2+3\lambda_b^2-3\lambda_\tau^2
+ \mbox{2-loop terms} \right]\,, \label{yuklam_b}
\end{equation}
with $d_i=(-4/3,0,16/3)$.  Evolution is mainly driven by the largest couplings
$g_3,\ \lambda_t,\ \lambda_b,\ \lambda_{\tau}$. The low-energy values at
$\mu=m_t$ are
\begin{eqnarray}\label{lambda_b}
\lambda_b(m_t) = {\sqrt2\, m_b(m_b)\over\eta_b v\cos\beta}\,, \qquad
\lambda_\tau(m_t) = {\sqrt2m_\tau(m_\tau)\over \eta_\tau v\cos\beta}\,, \qquad
\lambda_t(m_t) = {\sqrt2 m_t(m_t)\over v\sin\beta} \;,
\end{eqnarray}
where $\eta_f = m_f(m_f)/m_f(m_t)$ gives the running of the masses below
$\mu=m_t$, obtained from 3-loop QCD and 1-loop QED evolution, for heavy flavors
$f=t,b,c,\tau$.  For light flavors $f=s,u,d,e,\mu$ we stop at $\mu=1$~GeV and
define  $\eta_f = m_f(1{\rm\ GeV})/m_f(m_t)$. The $\eta_q$
values depend principally on the value of $\alpha_3(M_Z)$; for
$\alpha_3(M_Z)=0.
118,\
\eta_b\simeq 1.5,\ \eta_c\simeq 2.1,\ \eta_s=\eta_u=\eta_d\simeq 2.4$.
The running mass values
are $m_b(m_b)=4.25\pm0.15$~GeV, $m_\tau(m_\tau)=1.777$~GeV, $m_c(m_c) \simeq
1.2{\rm\ GeV}, m_s(1\rm\ GeV) \simeq 0.175$~GeV, $m_u(1\rm\ GeV) \simeq
0.006$~G
eV,
$m_d(1\rm\ GeV) \simeq 0.008$~GeV\cite{GL}.  The denominator factors in
Eq.~(\ref{lambda_b}) arise in the MSSM from the two Higgs vevs $v_1=v\cos\beta$
and $v_2=v\sin\beta$; they are related to the SM vev $v=246$~GeV by
$v_1^2+v_2^2=v^2$, while $\tan\beta=v_2/v_1$ measures their ratio.

\begin{center}
\hspace{0in}

{\small Fig.~2: Typical Yukawa scaling factors $S_i=
          \lambda_i(M_G)/\lambda_i(m_t)$~\cite{bbo}.}
\end{center}

The above RGE allow the Yukawa couplings to be
evolved from $m_t$ up to the GUT scale. Figure~2 illustrates
the scaling factors $S_i = \lambda_i(M_G)/\lambda_i(m_t)$ for the
case $M_{\rm SUSY} = m_t(m_t) =150$~GeV and $\alpha_s(M_Z)=0.118$.
It shows that the scaling factors are sensitive to $\tan\beta$
 when the latter is very large or very small. Large and small $\tan\beta$
correspond to the regions in which the Yukawa couplings become large; between
these regions and if $m_t$ is not too large ($\alt 170$ GeV), the gauge
couplings dominate the evolution giving rise to scaling factors less than one.

Figure 3 compares the values of $\lambda_i(M_G)$ corresponding to the scaling
factors in Figure 2.
({\em caution:} the input $b,c,s,d,u$ mass values here have substantial
 uncertainties).  Extrapolations of this kind allow us to
test various postulated GUT relations such as\cite{fixpt}
\begin{mathletters}
\begin{eqnarray}
\lambda_b(M_G) &\simeq& \phantom3\lambda_\tau(M_G)\,,~~\rm Ref.\cite{ceg} \;,
\label{b=tau} \\
\lambda_{\mu}(M_G) &\simeq& 3 \lambda_s(M_G)\,,~~\rm Ref.\cite{gj} \,, \\
 \lambda_e(M_G) &\simeq& \textstyle{1\over3}\lambda_d(M_G) \,,~~\rm
Ref.\cite{gj}\,.\\
\lambda_t(M_G) &\simeq& \lambda_b(M_G)
\simeq \lambda_{\tau}(M_G)\,,\label{tbteq}\rm~~
                                   Ref.\cite{tbt} \,.
\end{eqnarray}
\end{mathletters}
In the rest of this section we discuss various aspects of Yukawa coupling
evolution.

\begin{center}
\hspace{0in}

{\small Fig.~3: Typical Yukawa couplings $\lambda_i(M_G)$ at the
            GUT scale \cite{bbo}.}
\end{center}

The theoretical requirement that Yukawa couplings
remain perturbative throughout their evolution up to $M_G$ places
constraints on $\tan\beta$. If we require that the ratio of
2-loop/1-loop contributions in the RGE remains less than 1/4,
then\cite{bbo,cpw,lp}
\begin{equation}
                 0.6 \alt \tan\beta \alt 65   \,.
\end{equation}
There is also an indirect perturbative constraint on the
input parameter $\alpha_3(M_Z)$, if we wish to have $b$-$\tau$
Yukawa unification (Eq.\ref{b=tau}).  Since the $g_3^2$ and $\lambda_t^2$
terms enter the $\lambda_b/\lambda_{\tau}$ RGE with opposite
signs, an increase in $g_3$ requires a compensating increase
in $\lambda_t$ to maintain unification; see Fig.~4.  To keep
$\lambda_t$ perturbative requires\cite{bbo,cpw,lp}
\begin{equation}
            \alpha_3(M_Z) \alt 0.13 \,.
\end{equation}

As $\mu\to m_t$, $\lambda_t$ rapidly approaches an infrared
fixed point~\cite{pendleton}
as shown in Figure 5.
An approximate fixed-point solution for $m_t$ is given by the vanishing of the
one-loop terms on the right of Eq.~(\ref{yuklam_t})
\begin{equation}
-\sum c_ig_i^2 + 6\lambda_t^2 + \lambda_b^2 = 0 \,.
\end{equation}
Neglecting $g_1,\, g_2$ and $\lambda_b$, $m_t$ is then predicted in terms of
$\alpha_s(m_t)$ and $\beta$:~\cite{fixpt,cpw,lp,fkm,dhr,bbhz}
\begin{eqnarray}
m_t(m_t) \approx \frac{4}{3}\sqrt{2\pi\alpha_3(m_t)}\,
\frac{v}{\sqrt2}\sin\beta
\approx (192\gev){\tan\beta\over\sqrt{1+\tan^2\beta}} \,. \label{mt(mt)}
\end{eqnarray}
Thus the scale of the top-quark mass is naturally large in SUSY-GUT models but
depends on $\tan\beta$. Note that the propagator-pole mass is related to this
running mass by
\begin{equation}
m_t({\rm pole}) = m_t(m_t)\left[1+{4\over3\pi}\alpha_3(m_t)+\cdots\right]\,.
\label{mtpole}
\end{equation}

An exact numerical solution for the relation between $m_t$ and $\tan\beta$,
obtained from the 2-loop RGEs for $\lambda_t$ and $\lambda_b/\lambda_\tau$,
with $\lambda_b(M_G)=\lambda_{\tau}(M_G)$ unification, is shown in Fig.~6
taking $M_{\rm SUSY}=m_t$. At large $\tan\beta$, $\lambda_b$ becomes large and
the above fixed-point solution no longer applies. In fact, the solutions become
non-perturbative at large $\tan\beta$; our perturbative requirement
(2-loop)/(1-loop)${}\leq 1/4$ leads to $\lambda_t(M_G)\leq3.3$,
$\lambda_b(M_G)\leq3.1$ and $\tan\beta\alt 65$. For most $m_t$ values there are
two possible solutions for $\tan\beta$; the
lower solution is controlled by the $\lambda_t$ fixed point, following
Eqs.~(\ref{mt(mt)}),(\ref{mtpole}):
\begin{equation}
\sin\beta\simeq m_t(\rm pole)/(200\gev) \,. \label{mt3}
\end{equation}
An upper limit $m_t(\rm pole) \alt200$~GeV is found with these RGE solutions.

\begin{center}
\hspace{0in}

{\small Fig.~4: Qualitative dependence of  $\lambda_t$ at the GUT scale on
$\alpha_3(M_Z)$~\cite{bbo}.}
\end{center}

\bigskip

\begin{center}
\hspace{0in}

{\small Fig.~5: The Yukawa coupling $\lambda_t$ approaches a fixed
point at the electroweak scale~\cite{fixpt}.}
\end{center}

\bigskip

\begin{center}
\hspace*{0in}

{\small Fig.~6. Contours of constant $m_b(m_b)$ in the
$\left(m_t(m_t),\tan\beta^{\vphantom1}\right)$ plane~\cite{bbo}.}
\end{center}


Figures 3 and 6 show that there is the possibility of
$\lambda _t=\lambda _b=\lambda _{\tau }$ unification at $M_G^{}$;
$m_t$ and $\tan \beta $ must then be large.

In the presence of GUT threshold corrections, there may effectively be
corrections to GUT relations like Eq.~9\cite{lp,hs,hrs}.
Figure 7 shows the effects of small
deviations of $\lambda _b/\lambda _{\tau }$ from unity at $\mu =M_G^{}$.
Large threshold corrections are only possible in the case that
$\lambda _b < \lambda _{\tau }$ due to the proximity of the Landau pole.

\begin{center}
\hspace{0in}

{\small Fig.~7: GUT threshold corrections to Yukawa coupling
unification~\cite{fixpt}.}
\end{center}

\section{Evolution of the CKM matrix}

The CKM matrix comes from the mismatch
between transformations that diagonalize the up-type and down-type
quark mass matrices, arising from the matrices of Yukawa couplings.
We can therefore define a running CKM
matrix, with its own RGE, by diagonalizing the running mass matrices.
The RGE become especially simple if we keep only the leading terms
in the experimentally observed mass and CKM hierarchies, i.e.\ if
we neglect $\lambda_c/\lambda_t,\ \lambda_u/\lambda_c,\
\lambda_s/\lambda_b,\ \lambda_d/\lambda_s,\ |V_{ub}|^2$ and
$|V_{cb}|^2$\cite{op,bbo2,bs}.
Then the only off-diagonal CKM elements that evolve
are those connected to the third generation, i.e.\ $V_{ub},\ V_{cb},\ V_{td},\
 V_{ts}$, and these all have the same RGE (to all loops):
\begin{equation}
      \frac{d|V_{Qq}|}{dt} =
       -\frac{|V_{Qq}|}{16\pi ^2}[\lambda_t^2 +\lambda_b^2 + \mbox{2-loop}] \,.
\end{equation}
 All other off-diagonal matrix elements have $d|V_{Qq}|/dt=0$,
while the diagonal elements remain $\simeq 1$ by unitarity.  Hence
the moduli of CKM elements have the scaling behaviour\cite{op,bbo2,bs}
\begin{equation}
     |V_{\rm CKM}|(\mu=M_G) =  \left(
\begin{array}{rrr}
|V_{ud}| & |V_{us}| & \sqrt S |V_{ub}|\\
|V_{cd}| & |V_{cs}| & \sqrt S |V_{cb}|\\
\sqrt S |V_{td}| & \sqrt S |V_{ts}| & |V_{tb}|
\end{array}\right)_{\mu=m_t}
\end{equation}
 where $S$ is a universal scaling factor. A small unitarity violation
here is of sub-leading order in the mass/CKM hierarchy. Similarly,
it can be shown that the rephase-invariant CP-violation parameter
 $J = \Im(V_{ud}V_{cs}V_{us}^*V_{cd}^*)$ scales as\cite{op,bbo2,bs}
\begin{equation}
             J(\mu=M_G) = S J(\mu=m_t) \,.
\end{equation}
Figure 8 shows how the universal scaling factor  depends on
$\tan\beta$ in typical cases.  This approximate scaling property
offers a quick and simple way to test GUT-scale hypotheses
about mass- and CKM-matrices.

\begin{center}
\hspace{0in}

{\small Fig.~8: Typical CKM scaling factor $\sqrt S$ versus
$\tan\beta$~\cite{bbo2}.}
\end{center}

  GUT-scale Yukawa hypotheses usually take the form of assumed
parametric forms, called ``textures"\llap, for the quark and lepton mass
matrices at $M_G$.  For example, the SU(5) SUSY-GUT model of Ref.\cite{gj}
postulates up-quark, down-quark and charged lepton mass matrices
(Yukawa coupling matrices) of the forms
\begin{equation}
{\bf U} = \left(\begin{array}{ccc}
0 & C & 0 \\
C & 0 & B \\
0 & B & A \end{array}\right) \qquad
{\bf D} = \left(\begin{array}{ccc}
0 & F & 0 \\
F' & E & 0 \\
0 & 0 & D \end{array}\right) \qquad
{\bf E } = \left(\begin{array}{ccc}
0 & F & 0 \\
F' & -3E & 0 \\
0 & 0 & D\end{array}\right) \,.
\end{equation}
These immediately imply the relation\cite{gj,hrr,fls}
\begin{equation}
        |V_{cb}| = \sqrt {\lambda_c / \lambda_t } \label{Vcb} \,,
\end{equation}
since $V_{cb}$ originates entirely from the {\bf U}-matrix, and also
\begin{equation}
 \lambda_{\tau} \simeq \lambda_b \,, \qquad
 \lambda_{\mu} = 3 \lambda_s  \,,  \qquad
  \lambda_e = {1\over3} \lambda_d \,,
  \qquad     ({\rm all\ at}\ \mu = M_G)\,.
\end{equation}
The DHR model~\cite{dhr,hrr} introduces changes by
putting
\begin{equation}
{\bf U}= \left(\begin{array}{ccc}
0&C&0\\ C&0&B\\ 0&B&A\end{array}\right) \qquad
{\bf D}= \left(\begin{array}{ccc}
0&Fe^{i\phi}&0\\ Fe^{-i\phi}&E&0\\ 0&0&D\end{array}\right) \qquad
{\bf E}= \left(\begin{array}{ccc}
0&F&0\\ F&-3E&0\\ 0&0&D\end{array}\right) ,
\end{equation}
when the left and right down quarks and charged leptons appear in the same
multiplet such as the $\bf 16$ of SO(10).
Other phases can be rotated away. The ADHRS models\cite{adhrs}
\begin{equation}
{\bf U}= \left(\begin{array}{ccc}
0&{1\over 27}C&0\\ {1\over 27}C&0&x_uB\\ 0&x_u^{\prime }B&A\end{array}\right)
 \quad
{\bf D}= \left(\begin{array}{ccc}
0&C&0\\ C&Ee^{i\phi }&x_dB\\ 0&x_d^{\prime }B&A\end{array}\right) \quad
{\bf E}= \left(\begin{array}{ccc}
0&C&0\\ C&3Ee^{i\phi }&x_eB\\ 0&x_e^{\prime }B&A\end{array}\right) ,
\end{equation}
have fewer zeros in {\bf D} and {\bf E}
but retain SO(10)-type relations
$|V_{cb}| = \chi \sqrt{\lambda_c/\lambda_t}$ at $\mu = M_G$, where
$\chi$ is a (discrete) Clebsch factor. These models are more predictive since
they relate the up-quark Yukawa matrix to the down-quark Yukawa matrix
resulting in two fewer continuous parameters.

In order to test the prediction $|V_{cb}| = \sqrt{\lambda_c
/\lambda_t}$ at $\mu = M_G$, we can proceed as follows.
\begin{enumerate}
\item[(i)] Start with low-energy input, $m_t(m_t), m_b(m_b), m_c(m_c)$;
\item[(ii)] evolve Yukawa couplings up to $\mu = M_G$;
\item[(iii)] when $\lambda_b(M_G) = \lambda_{\tau}(M_G)$ is satisfied,
 construct $|V_{cb}(M_G)|$  and evolve it down to $|V_{cb}(m_t)|$, and
compare with experiment.
\end{enumerate}

Figure 9 shows contours of $|V_{cb}(m_t)|$ in the $(m_t(m_t),
\tan\beta)$ plane, for MSSM GUT solutions with the $b$-$\tau$
Yukawa unification constraint and various $m_c(m_c)$
input choices.  The region of $b$-$\tau$-$t$ Yukawa unification
is indicated. The relation Eq.~(\ref{Vcb}) leads to a lower bound\cite{bbo}
\begin{equation}
|V_{cb}(m_t)| \ge 0.043(200\ {\rm GeV}/m_t^{\rm pole})^{1/2}\;.
\end{equation}

\begin{center}
\hspace{0in}

{\small Fig.~9: Typical contours of $|V_{cb}(m_t)|$~\cite{bbo}.}
\end{center}

Figure 10 shows the effects of threshold corrections and/or a group theory
Clebsch factor on the GUT scale relation. These effects are parametrized
by the factor $X$ in the revised unification criteria
$|V_{cb}|=X\sqrt{\lambda _c/\lambda _t}$ at $M_G^{}$.

\begin{center}
\hspace{0in}

\parbox{5.5in}{\small Fig.~10: Effects of threshold corrections or a Clebsch
factor\cite{adhrs}
on $|V_{cb}|=\sqrt{\lambda _c/\lambda _t}$ unification~\cite{bbo}.}
\end{center}

\bigskip

Some relations between quark masses and CKM mixing angles are satisfied
under more general assumptions. For example, the GUT scale relationships
$|V_{ub}/V_{cb}|=\sqrt{m_u/m_c}$ and $|V_{td}/V_{ud}|=\sqrt{m_d/m_s}$
have been shown to
pertain to a whole class of unification scenarios\cite{hr}.

\section{Implications for SUSY Higgs searches}

   The $\lambda_t$ fixed-point solutions have interesting implications
for the phenomenology of Higgs bosons in the MSSM.
Recall that there are 5 Higgs bosons in this model\cite{hhg}:
neutral CP-even $h$ and $H\ (m_h < m_H)$, neutral CP-odd $A$, and charged
$H^{\pm}$.
At tree level there are just two parameters, usually taken to be $m_A$ and
$\tan\beta$ and a mass bound $m_h < M_Z$, but large one-loop radiative
corrections (principally depending on $m_t$ and the mean $t$-squark mass
$m_{\tilde t}$) affect the Higgs masses and couplings and push up the $m_h$
bound to
\begin{equation}
m_h^2 < M_Z^2 + \frac{6G_Fm_t^4}{\sqrt2 \pi^2} \ln(m_{\tilde t}/m_t) .
\end{equation}
Studies of MSSM Higgses usually refer to the $(m_A, \tan\beta)$ parameter
plane, taking a range of $m_t$ and assuming $m_{\tilde t} \sim 1$~TeV.

   Several groups\cite{barger,baer,gunion,kunszt} have systematically
discussed the potential of present and future colliders to discover one
or more MSSM Higgses.  LEP\,I has already excluded part of the $(m_A,
\tan\beta)$ plane; LEP\,II will cover more but not all. SSC/LHC offer new
search possibilities, but there remains a region $m_A \sim 100$--150~GeV,
$\tan\beta \agt 5$  where apparently no MSSM Higgs signals whatever would
be detectable (unless high-performance $b$-tagging\cite{dai} and rapidity gap
searches\cite{fs} can succeed).

   It is therefore very interesting to find any theoretical arguments why
this inaccessible region may be forbidden.  The $\lambda_t$ fixed-point
solutions provide a possible argument if $m_t \alt 160$~GeV, since
these solutions are then constrained to a range of small $\tan\beta$ as
shown in Fig.~6, with $m_t$ and $\tan\beta$ directly correlated via
Eq.~(\ref{mt3})~\cite{fixpt}.  With this correlation, the $(m_A,\tan\beta)$
region excluded by 1992 LEP\,I searches is shown in Fig.~11(a) (assuming
$m_{\tilde t} \simeq 1$~TeV).  The corresponding region in the
$(m_h,\tan\beta)$ plane
is shown in Fig.~11(b), where the left-hand boundary comes from LEP\,I data
and the right-hand boundary comes from internal MSSM constraints with
one-loop corrections.  We see that the lower limit on $m_h$ is about 60
GeV (close to the 61~GeV SM Higgs limit with these data), while the input
assumption $m_t < 160$~GeV implies that $m_h \alt 85 $~GeV --- within
reach of LEP\,II searches.  The corresponding mass limits on the other
Higgses are $m_A \agt 70$~GeV, $m_{H^\pm} \agt 105$~GeV, $m_H \agt 140$~GeV.
In principle, $A$ too might be discoverable at LEP\,II via $e^+ e^- \to Ah$
production, but in fact there is only a small parameter region where
the cross section would be big enough; the other production channels
$AH,ZH,H^+H^-$ are kinematically inaccessible.

\begin{center}
\hspace{0in}

\parbox{5.5in}{\small Fig.~11: $\lambda_t$-fixed-point solution regions allowed
by the LEP\,I data: (a)~in the $(m_A, \tan \beta )$ plane, (b)~in the $(m_h,
\tan \beta )$ plane. The top quark masses are $m_t({\rm pole})$, correlated to
$\tan \beta $ by Eq.~(\ref{mt3})~\cite{fixpt}.}
\end{center}

   Thus this range of $\lambda_t$ fixed-point solutions implies that we
shall not have to wait for SSC/LHC to discover a MSSM Higgs boson.  What
more will be detectable there?  Figure~12 shows the limits of
detectability for the principal SSC/LHC signals, in the $h \to \gamma
\gamma,\  H \to \ell\ell\ell\ell, A \to \gamma\gamma$  and $H^{\pm} \to
\tau\nu$ channels.  Depending on $m_A$ and $\tan\beta$, we see there could be
good chances to discover one or more additional Higgses, though not all
of them at once; but there also exists a parameter region where no Higgs
signals whatever would be expected at SSC/LHC\cite{fixpt}.

\begin{center}
\hspace{0in}

\parbox{5.5in}{\small Fig.~12: SSC/LHC signal detectability regions, compared
with the LEP\,I allowed region of $\lambda_t$-fixed-point solutions from
Fig.~11(a) and the  probable reach of LEP\,II.
   The top quark masses are $m_t({\rm pole})$~\cite{fixpt}.}
\end{center}

   Possible future $e^+e^-$ linear colliders with energies above LEP\,II
offer interesting further possibilities, however.  The principal
neutral-Higgs production channels are
\begin{eqnarray}
  e^+ e^- &\to& Zh, Ah, ZH, AH
\\
  e^+ e^- &\to& \nu \nu h, \nu \nu H, e^+ e^- h, e^+ e^- H .
\end{eqnarray}
Here the two-body cross sections fall with $1/s$ while the others
($WW$ and $ZZ$ fusion) rise logarithmically.  Now the $Z^* \to ZH,Ah$ plus
$WW,ZZ \to H$ rates are all suppressed by a factor $\cos^2(\beta-\alpha)$,
where
$\alpha$ is a CP-even mixing angle; in the $\lambda_t$ fixed-point solutions,
$\cos^2(\beta-\alpha) < 0.3\ (0.05)$ for $m_t < 160$~GeV (145~GeV).  However,
the remaining $Z^* \to Zh,AH$ plus $WW,ZZ \to h$ rates
contain the complementary factor $\sin^2(\beta - \alpha)$ and are
unsuppressed, while the charged-Higgs process
\begin{equation}
  e^+ e^- \to H^+ H^-
\end{equation}
has no such factors.  Copious $h$ production is therefore guaranteed, with
$H, A, H^{\pm}$ too if they are not too heavy.

\section{Summary}
\begin{enumerate}
\item[a)] The success of SUSY GUTS in gauge coupling unification is
tantalizing.
\item[b)] Yukawa coupling possibilities $(\lambda_b \simeq \lambda_{\tau}$,
etc) are equally attractive.
\item[c)] The constraint $\lambda_b(M_G) = \lambda_{\tau}(M_G)$ leads to a
narrow corridor in the plane of $\tan\beta$ and $m_t^{\rm pole}$.
\item[d)] $\lambda_t$ fixed-point solutions with $\alpha_s(M_Z) = 0.118$
predict $\sin\beta \simeq m_t^{\rm pole}/(200$~GeV) or $\tan \beta $
large.
\item[e)] Perturbativity at the GUT scale implies several constraints:
   $m_t^{\rm pole} \alt 200$~GeV (for $\alpha_s(M_Z) = 0.118$),
   $\alpha_s(M_Z) \alt 0.13$,
   $\tan\beta \alt 65$.
\item[f)] A simple scaling law connects CKM matrix elements at $\mu = m_t$ and
$M_G$.
\item[g)] GUT textures give interesting low-energy predictions; e.g.
   $|V_{cb}(M_G)| = \hfill\break
 \sqrt {\lambda_c(M_G)/\lambda_t(M_G)}$ gives
   $|V_{cb}(m_t)| > 0.043 (200\ {\rm GeV}/m_t^{\rm pole})^{1/2}$.
\item[h)] Threshold effects at the GUT scale may not be negligible.
\item[i)] $\lambda_t$ fixed-point solutions imply that $m_t \agt 130$~GeV and
   the lightest MSSM Higgs mass $m_h \agt 60 GeV$; if in fact $m_t \alt
 160$~GeV, then $m_h \alt 85$~GeV and $h$ will be discoverable at LEP.
\end{enumerate}

{\center{ACKNOWLEDGEMENTS}}
This research was supported
in part by the University of Wisconsin Research Committee with funds granted by
the Wisconsin Alumni Research Foundation, in part by the U.S.~Department of
Energy under contract no.~DE-AC02-76ER00881, and in part by the Texas National
Laboratory Research Commission under grant no.~RGFY93-221. PO was supported in
part by an NSF Graduate Fellowship.


\begin{thebibliography}{00}
\frenchspacing

\bibitem{giatw} Review of Particle Properties, Phys. Rev. {\bf D45,} no.11-II
(1992); G.~Altarelli, CERN-TH.6623/92.

\bibitem{susygut} U.~Amaldi, W.~de~Boer, and H.~Furstenau,
Phys.\ Lett.\ {\bf B260}, 447 (1991); J.~Ellis, S.~Kelley and
D.~V.~Nanopoulos, Phys.\ Lett.\ {\bf B260} 131 (1991);
P.~Langacker and M.~Luo, Phys.\ Rev.\ {\bf D44}, 817 (1991).

\bibitem{bbo}V.~Barger, M.S.~Berger, and P.~Ohmann, Phys.\ Rev.\ {\bf D47},
1093 (1993), and unpublished calculations.

\bibitem{susyrge1} K.~Inoue, A.~Kakuto, H.~Komatsu and S.~Takeshita,
Prog.\ Theor.\ Phys.\ {\bf 67}, 1889 (1982).

\bibitem{susyrge2} J.~E.~Bj\"{o}rkman and D.~R.~T.~Jones, Nucl.\ Phys.\ {\bf
B25
9}, 533 (1985).

\bibitem{GL} J.~Gasser and H.~Leutwyler, Phys.\ Rep.\ {\bf 87}, 77 (1982).

\bibitem{fixpt}V.~Barger et al., Madison preprint MAD/PH/755, to be
  published in Phys. Lett.~B.

\bibitem{ceg}M.~Chanowitz, J.~Ellis, and M.~K.~Gaillard, Nucl.\ Phys.\
{\bf B128}, 506 (1977).

\bibitem{gj}H.~Georgi and C.~Jarlskog, Phys. Lett. {\bf 86B}, 297 (1979).

\bibitem{tbt} M.~Olechowski and S.~Pokorski, Phys. Lett. {\bf B214}, 393
(1991);
B.~Ananthanarayan et al., Phys. Rev. {\bf D44}, 1613 (1991);
S.~Kelley et al., Phys. Lett. {\bf B274}, 387 (1992);
B.~Ananthanarayan, G.~Lazarides, and Q.~Shafi, Phys.\ Lett.\ {\bf B300},
245 (1993).

\bibitem{cpw} M.~Carena, S.~Pokorski, and C.~E.~M.~Wagner, Max Planck Institute
preprint MPI-PH-93-10 (1993).

\bibitem{lp} P.~Langacker and N.~Polonsky, University of Pennsylvania preprint
UPR-0556-T (1993).

\bibitem{pendleton} B.~Pendleton and G.~G.~Ross, Phys.\ Lett.\ {\bf 98B}, 291
(1981); C.~T.~Hill, Phys.\ Rev.\ {\bf D24}, 691 (1981).

\bibitem{fkm} C.~D.~Froggatt, I.~G.~Knowles, and R.~G.~Moorehouse, Phys.\
Lett.\ {\bf B249}, 273 (1990); {\bf B298}, 356 (1993).

\bibitem{dhr} S.~Dimopoulos, L.~Hall and S.~Raby, Phys.\ Rev.\ Lett. {\bf68},
1984 (1992); Phys.\ Rev.\ {\bf D45}, 4192 (1992); {\bf D46}, 4793 (1992);
G.W.~Anderson et al., Phys. Rev. {\bf D47}, 3702 (1993).

\bibitem{bbhz} V.~Barger, M.~S.~Berger, T.~Han, and M.~Zralek, Phys.\ Rev.\
Lett.\ {\bf68}, 3394 (1992).

\bibitem{hs} A.~E.~Faraggi, B.~Grinstein, S.~Meshkov, SSCL-PREPRINT-126-REV;
L.~J.~Hall and U.~Sarid, Lawrence Berkeley Preprint LBL-32905
(1993).

\bibitem{hrs} L.~J.~Hall, R.~Rattazzi, and U.~Sarid, Lawrence Berkeley Preprint
LBL-33997 (1993).

\bibitem{op} M.~Olechowski and S.~Pokorski, Phys. Lett. {\bf B257}, 388 (1991).

\bibitem{bbo2}V.~Barger, M.S.~Berger and P.~Ohmann, Phys. Rev. {\bf D47}, 2038
(1993).

\bibitem{bs} K.~S.~Babu and Q.~Shafi, Phys.\ Rev.\ {\bf D47}, 5004 (1993).

\bibitem{hrr} H.~Georgi and D.~Nanopoulos, Nucl.\ Phys.\ {\bf B159}, 16 (1979);
J.~Harvey, P.~Ramond, and D.~B.~Reiss, Phys.\ Lett.\ {\bf92B},
309 (1980); Nucl.\ Phys.\ {\bf B199}, 223 (1982).

\bibitem{fls} E.~M.~Freire, G.~Lazarides, and Q.~Shafi, Mod.\ Phys.\
{\bf A5}, 2453 (1990).

\bibitem{adhrs} L.~J.~Hall, Plenary talk given at 16th Texas Symposium  on
Relativistic Astrophysics and 3rd Particles, Strings, and Cosmology Symposium
(TEXAS / PASCOS 92), Berkeley, CA, 13-18 Dec 1992, LBL-33677 (1993);
S.~Raby, Talk presented at International Workshop on Recent Advances in the
Superworld, Woodlands, TX, 13-16 Apr 1993, OHSTPY-HEP-T-93007 (1993).

\bibitem{hr} L.~J.~Hall and A.~Rasin, Lawrence Berkeley Laboratory preprint
LBL-33668 (1993).

\bibitem{hhg} For a review see J.F.~Gunion, H.E.~Haber, G.L.~Kane and
S.~Dawson, ``The Higgs-Hunter's Guide", Addison-Wesley (1990).

\bibitem{barger}V.~Barger et al., Phys. Rev. {\bf D45}, 4128; {\bf D46}, 4914
(1992).

\bibitem{baer}H.~Baer et al., Phys. Rev. {\bf D46}, 1067 (1992).

\bibitem{gunion}J.F.~Gunion et al., Phys. Rev. {\bf D46}, 2040, 2052 (1992);
  {\bf D47}, 1030 (1993).

\bibitem{kunszt}Z.~Kunszt and F.~Zwirner, Nucl. Phys. {\bf B385}, 3 (1992).

\bibitem{dai}J.~Dai, J.F.~Gunion and R.~Vega, Davis preprint UCD-93-20;
T.~Garavaglia, W.~Kwong, and D.~Wu, Prairie View A-M Preprint PVAM-HEP-93-1.

\bibitem{fs} R.~S.~Fletcher and T.~Stelzer, Madison preprint MAD/PH/763 (1993),
and unpublished calculations.

\end{thebibliography}
\end{document}